\documentstyle[psfig,preprint,aps]{revtex}

\tightenlines

\newcommand{\sla}{\!\!\!\!/ \,}

\def\beq{\begin{equation}}
\def\eeq{\end{equation}}
\def\bea{\begin{eqnarray}}
\def\eea{\end{eqnarray}}

\begin{document}
\draft

\title{Quark Dispersion Relation and Dilepton Production in the
Quark-Gluon Plasma}

\author{Andr\'e Peshier$^1$ and Markus H. Thoma$^2$}
\address{$^1$Forschungszentrum Rossendorf, PF 510119, 01314 Dresden, Germany}
\address{$^2$Institut f\"ur Theoretische Physik, Universit\"at Giessen, 35392
Giessen, Germany}

\maketitle

\begin{abstract}
Under very general assumptions we show that the quark dispersion relation in
the quark-gluon plasma is given by two collective branches, of which one has
a minimum at a non-vanishing momentum. This general feature of the quark
dispersion relation leads to structures (van Hove singularities, gaps) in the
low mass dilepton production rate, which might provide a unique signature for
the quark-gluon plasma formation in relativistic heavy ion collisions.
\end{abstract}

\bigskip

\pacs{PACS numbers: 12.38.Mh, 25.75.-q}

\newpage

Relativistic heavy ion experiments at SPS, RHIC, and LHC are and will be
undertaken in order to produce a new state of matter, the so-called
quark-gluon plasma (QGP). This deconfined phase of quarks and gluons in or
close to thermal equilibrium might be present in the hot and dense fireball
of such nucleus-nucleus collisions for about 10 fm/c \cite{ref1}.
The biggest problem in the discovery of the QGP is to find a clear signature
for its formation \cite{ref2}. Hadrons arriving in the detectors carry mainly
information about the hadronic phase following the QGP in the expansion of
the fireball. Thermal photons, on the other hand, are emitted from the
fireball without any further interaction \cite{ref3}. Therefore they are a
unique tool to test the QGP phase. In particular, virtual photons, which
decay into lepton pairs ($e^+e^-$, $\mu^+\mu^-$), could serve as a promising
signal for the QGP formation \cite{ref4}. For this purpose one has to
calculate the dilepton spectrum from a fireball with and without phase
transition and to compare it with the experiments. Unfortunately, the
dilepton production rate from the QGP seems to be similar to the one from
a hadron gas at the same temperature \cite{ref5}. However, this observation
(duality) is based on the assumption that the dilepton production in the QGP
arises solely from annihilation of bare quarks (Born term) \cite{ref6}.
As we will argue in the following, medium effects will change the quark
dispersion relation in the QGP in a way that sharp structures (singularities,
gaps) arise in the production rate of low mass dileptons. These structures
might provide a clear and unique signature for the presence of deconfined,
collective quarks in the fireball.

The starting point of our investigation is the most general expression
for the self energy of fermions in the chiral limit of vanishing current
mass \cite{foot1},
which in the rest frame of the medium has the form \cite{ref7}
\beq
\Sigma(P)
 =
 -a P\sla - b\gamma_0
 =
 -(ap^0+b)\gamma_0 + a\mbox{\boldmath$p$} \cdot \mbox{\boldmath$\gamma$} \, .
\label{se}
\eeq
The scalar quantities $a$ and $b$ are functions of the energy $p^0$ and
the magnitude $p$ of the three momentum, with certain properties to be
studied below. The fermion dispersion relation is given by the position
of the poles of the exact propagator $S = (S_0^{-1}-\Sigma)^{-1}$.
Decomposed into helicity eigenstates, with $\hat{\mbox{\boldmath$p$}} =
\mbox{\boldmath$p$} / p$, the propagator reads \cite{ref8}
\beq
S(P)
=
\frac{\gamma_0-\hat{\mbox{\boldmath$p$}}\cdot \mbox{\boldmath$\gamma$}}{2D_+(P)}
+\frac{\gamma_0+\hat{\mbox{\boldmath$p$}}\cdot \mbox{\boldmath$\gamma$}}{2D_-(P)} \, .
\label{prop}
\eeq
The zeros $\omega_\pm(p)$ of
\beq
D_\pm = -p^0 + T_0 \pm (1+a)p \, , \quad T_0 = -(ap^0+b) \, ,
\label{zero}
\eeq
describe the propagation of particle excitations $q_+$ with energy
$\omega_+$ and of a mode $q_-$ called plasmino \cite{ref8} with
energy $\omega_-$ and negative ratio of chirality to helicity. The
latter is a consequence of the medium, breaking the Lorentz invariance
of the vacuum. Invariance under charge conjugation implies that
$-\omega_\pm (p)$ are also solutions of (\ref{zero}) \cite{Weldon}.
In the following, we focus on the solutions with positive energy.

Considering quark excitations in the QGP, we emphasize the existence of a
real part of the solutions of the exact (implicit) dispersion relations
\beq
\omega_\pm(p) = T_0 \pm (1+a)p
\label{omega}
\eeq
according to the assumption of a deconfined phase. Furthermore we note that
at large momentum the breaking of the Lorentz symmetry is less manifest,
hence $|b|\ll |ap^0|$ in the self energy (\ref{se}), and the dispersion
relations are expected to approach the light cone.
While this is evident for the particle branch $\omega_+$, we speculate that
this behavior also holds for the plasmino branch.
In the two different approximations for the in-medium quark propagator,
discussed below, the dispersion relations (\ref{omega}) in the form
$\omega _\pm=f(\omega _\pm,p)\pm p$ are determined by the function
$f=-b/(1+a)$, having a space-like pole close to the light cone.
Taking this as a general feature, the plasmino branch indeed approaches 
the light cone for large momenta as well.

At vanishing $p$, due to isotropy, the self energy (\ref{se}) is independent
of the direction $\hat{\mbox{\boldmath$p$}}$ of the momentum, thus $a p 
\rightarrow 0$ for $p \rightarrow 0$.
Consequently, both excitations have the same rest energy \cite{ref9}, which
we will refer to as an effective mass, $\omega_\pm(0)=m_q^*$.
According to (\ref{omega}), the slopes of the branches of the dispersion
relation are implicitly given by
\beq
\omega_\pm' = \dot{T}_0\,\omega_\pm' + T_0' \pm
\left( 1+a+(\dot{a}\omega_\pm'+a')\,p \right) \, ,
\label{omega'}
\eeq
where the derivatives $\partial/\partial p^0$ and $\partial/\partial p$ were
denoted by dots and primes, respectively. As we will show in the following,
several terms in (\ref{omega'}) vanish for small momenta due to isotropy
properties of Green's functions, similar to the argument applied above for
the self energy. Using the differential Ward identity in a ghost free
gauge \cite{foot2}, the quark-gluon vertex is related to the quark
propagator by $\Gamma_\mu(P,P'\rightarrow P) = \partial S^{-1} / \partial
P^\mu$ \cite{ref11}. The isotropy of the temporal and spatial components of
$\Gamma_\mu$ at vanishing momentum results from
\beq
\dot{a}p \rightarrow 0 \, , \; a'p^0+b' \rightarrow 0 \, , \;
a'p      \rightarrow 0    \quad \mbox{for } p \rightarrow 0 \, ,
\label{regprop}
\eeq
which leads to
\beq
\omega_\pm'(0) = \pm \frac{1+a(p_0=m_q^*,p=0)}{1-\dot{T}_0(p_0=m_q^*, p=0)}
\, .
\label{omega0'}
\eeq
Evidently, the initial slopes of both branches are opposite \cite{Weldon_new}
and, as can also be derived from the regularity properties (\ref{regprop}),
neither vanishing nor infinite. From (\ref{omega0'}) we conclude that one
branch of the chiral quark dispersion relation (the plasmino mode, as it turns
out) has a minimum before it approaches the light cone.

This behavior has been found in perturbative as well as non-perturbative
approximations of the quark propagator. In the perturbative calculation
\cite{ref8} (hard-thermal-loop approximation) the effective quark mass is
given by $m_q^*=gT/\sqrt{6}$ and the plasmino branch shows a minimum at
$p_{min}=0.408 m_q^*$ before approaching the light cone. The spectral
strength of the plasmino branch (i.\,e., the residue of the pole of the 
propagator) vanishes exponentially close to the light cone, which shows
that plasminos are purely collective excitations.
In the case of a finite current quark mass there is a splitting of the
two dispersion relations at zero momentum and the minimum of the
plasmino branch vanishes for large current masses \cite{blaizot}.
Similar dispersion relations were found in a non-perturbative calculation
\cite{ref12} of the quark propagator based on the presence of a gluon
condensate, which has been measured in lattice QCD simulations, in the QGP.
They differ from the perturbative results by an effective quark mass
$m_q^*=1.15 T$ between the critical temperature $T_c$ and $4T_c$ and
a powerlike vanishing spectral strength of the plasmino mode close to
the light cone. As an example, the quark dispersion relation following from
the non-perturbative approach is shown at $T=2T_c$ in Fig.\,1.

Now we will discuss the consequence of this quark dispersion relation on the
production of lepton pairs from the QGP. Virtual photons with invariant mass
$M=\sqrt{k_0^2-k^2}$ are emitted either by an electromagnetic transition from
the upper to the lower branch, $q_+\rightarrow q_-\gamma^*$, or by
annihilation, e.\,g. $q_-\bar q_-\rightarrow \gamma^*$.
At vanishing momentum of the virtual photon, $k=0$, the first process yields
a contribution to the dilepton production rate, which starts at $M=0$ and
terminates at the maximum difference of $E(p)=\omega _+(p)-\omega _-(p)$
of the two branches. There we encounter a van Hove singularity which is caused
by the divergence of the density of states, which is proportional to
$(dE(p)/dp)^{-1}$ \cite{ref13}.
For larger $M$, there is a gap in the dilepton rate before the channel
from plasmino annihilation opens at the threshold $M=2\omega _-(p_{min})$
with another van Hove singularity originating from the minimum in the plasmino
dispersion. At $M=2m_q^*$, the annihilation of collective quarks,
$q_+\bar q_+\rightarrow \gamma^*$, sets in. This contribution dominates at
large invariant mass, where the plasmino contribution is suppressed, and
finally approaches the Born term.

As an illustration, the dilepton rate following from the imaginary part
of the photon self energy using the effective quark propagators that
contain the gluon condensate as calculated in \cite{ref12}, corresponding
to the dispersion relation of Fig.\,1, is shown for $T=2T_c$ in Fig.\,2.
The same structures of the spectrum are observed in perturbative
calculations based on the hard-thermal-loop resummation
technique \cite{ref8,foot3}.

The validity of perturbative calculations as \cite{ref8} assuming $g \ll 1$ is
doubtful at temperatures within reach of the experiments and the reliability
of phenomenological non-perturbative estimates, as the one discussed in
\cite{ref12}, is difficult to control. Fortunately, the appearance of a
distinct structure in the dilepton rate is a general feature independent
of the approximation assumed for the quark propagator, as we have argued.
The effective quark mass $m_q^*$ is expected to be of the order of 0.5 GeV in
the temperature regime under consideration, as can be deduced by comparing
lattice calculations for the equation of state of the QGP with an ideal gas
model of quasiparticles, i.\,e., quarks and gluons with effective temperature
dependent masses \cite{ref14}. Therefore we predict that these structures
manifest themselves in the dilepton spectrum for invariant masses of around
1 GeV and below.

If, contrary to the general expectation, the plasmino branch were a
monotonically decreasing function of the momentum, which intersects
the light cone, van Hove singularities in the dilepton rate would be
absent. Nevertheless, even in this case there would be a significant
enhancement of the rate at small momentum compared to the Born rate
due to the transition $q_+\rightarrow q_-\gamma^*$.
The behavior of the plasmino branch, however, can possibly be studied
in lattice simulations of the quark propagator at finite temperature
\cite{karsch}. The anticipated existence of the plasmino minimum could,
hence, be corroborated by such ab-initio calculations which also allow
to consider nonzero current quark masses.

Whether the structures in the dilepton spectrum can be experimentally observed
or not is a difficult question. Definitely they will be smeared out and, at
least partially, covered by various effects. First of all, there will be smooth
contributions to the dilepton rate such as Compton scattering, $q\bar q$
annihilation with gluon emission, and bremsstrahlung, which might cover up
the van Hove singularities. In the hard-thermal-loop calculation these
processes are represented by cut contributions of the photon self energy,
which are due to the imaginary part of the hard-thermal-loop quark self energy
entering the resummed quark propagator. In the gluon condensate calculation,
on the other hand, cut contributions are not present due to the fact that the 
quark self energy is real in this case. 
In our general consideration, however, for covering up the van Hove peaks
by smooth higher order contributions completely, these contributions also 
had to dominate clearly over the Born term and therefore, as well, over the
hadronic contributions according to Ref.\cite{ref5}. If this is not ruled 
out already by SPS data \cite{ceres}, a strong enhancement of the low mass 
dilepton rate caused by cut and higher order QCD contributions could serve 
itself as a signal for the QGP formation.

Furthermore, the sharp structures will be smeared out by contributions to
the dilepton spectrum at finite photon momentum \cite{ref15} and due to 
damping effects, which we have not touched in our discussion of the real  
parts of the poles. Another smoothing of these structures originates from 
the space-time evolution of the fireball, with which the rate has to be
convoluted in order to extract the dilepton spectrum. To what extent
the sharp structures in the dilepton rate, coming from the in-medium
quark dispersion relation, will survive is an open question.
After all it will be worthwhile looking for new structures in the spectrum
of low mass dileptons with small transverse momentum. Such structures could
not be seen at SPS \cite{ceres}
because the contributions from the QGP phase to the
dilepton spectrum are estimated by hydrodynamical calculations
\cite{ref17} to be one or two orders of magnitude below the data due to
the small life time of the QGP.
At RHIC and LHC, however, where the QGP phase is expected to
dominate the dilepton spectrum, these structures could show up.

Summarizing, we have argued that in general quarks in the QGP possess a
dispersion relation corresponding, at small momentum, to two massive
collective modes, and that one branch has a minimum at finite momentum.
As a consequence of this general feature, the dilepton production rate
exhibits sharp structures (peaks and  gaps) at invariant masses below
about 1 GeV. Dilepton rates from a hadron gas, on the other hand, seem to
be rather flat due to collisional broadening of the resonances \cite{ref5}.
Therefore the observation of new structures in the low mass dilepton
spectrum at RHIC and LHC would be a strong indication for the presence of
collective excitations of deconfined quarks in the fireball. But even if
van Hove peaks cannot be seen, we expect in the case of a QGP
formation a significant enhancement of the low mass dilepton rate due to 
the existence of a plasmino branch and due to higher order contributions 
compared to the hadronic predictions \cite{ref5}. In conclusion, we predict
that there will be a non-trivial QGP contribution to the low mass dilepton
spectrum at RHIC and LHC and therefore the dilepton spectrum is a
promising candidate for revealing the QGP phase.

\vspace*{0.5cm}

\centerline{\bf ACKNOWLEDGMENTS}
\vspace*{0.3cm}
We would like to thank C. Greiner, V. Koch, S. Leupold, U. Mosel,
M. Mustafa, and A. Weldon for stimulating and helpful discussions.
This work was supported by BMBF, GSI Darmstadt, and DFG.

\vspace{-0.5cm}

\begin{figure}
\centerline{\psfig{figure=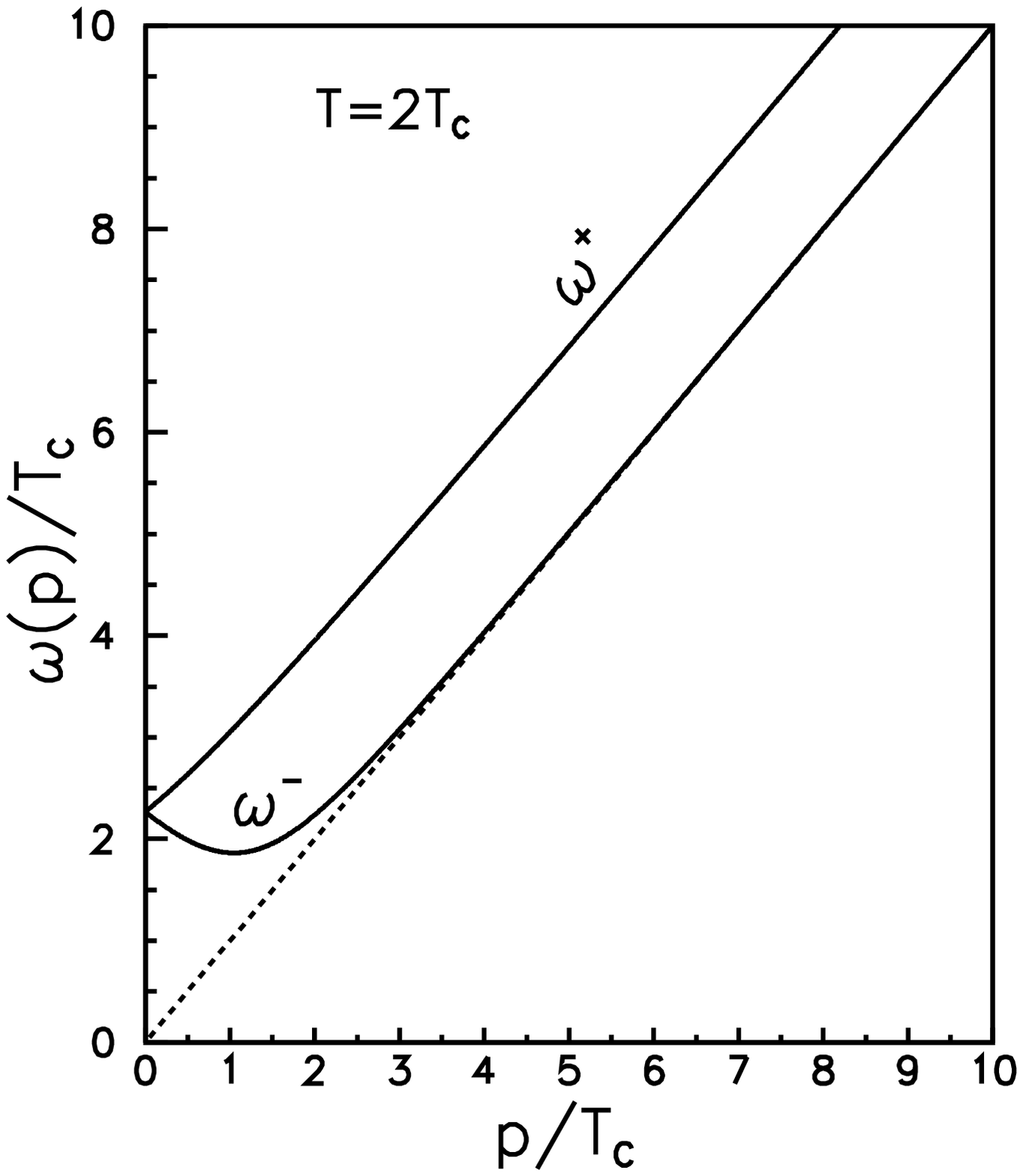,width=9cm}}
\vspace*{-1.5cm}
\caption{Example for the quark dispersion relation in the QGP based on
a quark propagator containing the gluon condensate [16].}
\end{figure}

\vspace{-0.5cm}

\begin{figure}
\centerline{\psfig{figure=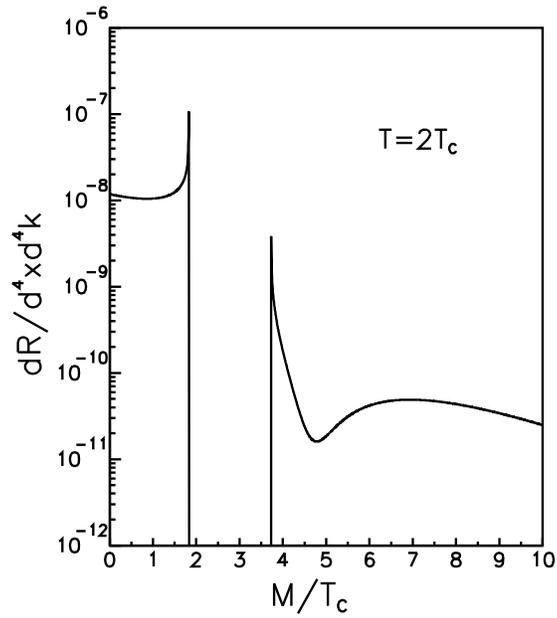,width=9cm}}
\vspace*{-1.5cm}
\caption{Example for the dilepton production rate at zero photon momentum
following from the quark dispersion of Fig.1 in the QGP [16].}
\end{figure}


\begin{references}
\bibitem{foot1} In the QGP, the masses of the most abundant light quark
flavors, up and down, can be neglected compared to their thermal energy.
\end{references}
\end{document}